\documentclass[12pt]{article}

\usepackage{graphicx}
\usepackage{amsmath}
\usepackage{amssymb}

\numberwithin{equation}{section}

\usepackage{color}
\input{colordvi.tex}

\allowdisplaybreaks

\newcommand{\nn}{\nonumber\\}

\newcommand{\vac}[1]{\left|{0}\right>}

\begin{document}

\begin{titlepage}
\thispagestyle{empty}
\begin{flushleft}
\hfill September, 2011
\end{flushleft}

\vskip 1.5 cm
\bigskip

\begin{center}
{\Large\bf{
Homotopy Operators and Identity-Based Solutions 
}}
\vskip 1em
{\Large\bf{
in Cubic Superstring Field Theory
}}\\

\renewcommand{\thefootnote}{\fnsymbol{footnote}}

\vskip 2.5cm
{\large
Shoko {\sc Inatomi}$^{1}$,
Isao {\sc Kishimoto}$^{2}$ 
and Tomohiko {\sc Takahashi}$^{1}$}\\

\bigskip\bigskip

{\it
$^{1)}$ Department of Physics, Nara Women's University,
Nara 630-8506, Japan,\\
$^{2}$ Faculty of Education, Niigata University,
Niigata 950-2181, Japan
}
\vskip 2in
\end{center}
\begin{abstract}
We construct a class of nilpotent operators
using the BRST current and ghost fields in superstring theory.
The operator can be realized  in cubic superstring field theory
as a kinetic operator in the background of
an identity-based solution.
For a particular type of the deformed BRST operators,
we find a homotopy operator and discuss its relationship to the
cohomology in a similar way to the bosonic case,
which has been elaborated by the authors.

\end{abstract}
\end{titlepage}\vfill\setcounter{footnote}{0} \renewcommand{\thefootnote}{
\arabic{footnote}} \newpage

\section{Introduction}

String field theory (SFT) has been used in exploring the non-perturbative
vacuum in string theory.
After the success of the Schnabl solution\,\cite{Schnabl:2005gv}, strategy of
constructing various classical solutions in SFT has been developed in
terms of ``$KBc$ subalgebra''\cite{Okawa:2006vm} and its extension.\footnote{
See \cite{Fuchs:2008cc} and references therein, for example.
}
In modified cubic superstring field theory, the Erler
solution\,\cite{Erler:2007xt}
was constructed as a straightforward extension of the Schnabl solution.
These solutions are impressive, in particular, in the sense that the
vacuum energy was evaluated exactly and turned out to be equal to 
the D-brane tension.
Furthermore, homotopy operators for the BRST operators at the solutions
have been constructed as in \cite{Ellwood:2006ba}, and it shows the
empty cohomology at all ghost number sectors.

On the other hand, another type of classical solutions starting from
\cite{Takahashi:2002ez} 
have been investigated, which are called a class of ``identity-based
solutions.''
They are constructed by half-integrations of the BRST current
and the $c$-ghost with particular functions on the identity state.
Although it is difficult to evaluate its vacuum energy directly
due to a singular property of the identity state,\footnote{
Numerical quantitative evidences have been obtained 
indirectly up to level 26~\cite{Kishimoto:2009nd,
Kishimoto:2009hc,Kishimoto:2011zz}.
}
the cohomology of the BRST operator of the theory around the solution
was investigated and turned out to be empty in the ghost number one
sector \cite{Kishimoto:2002xi}, which provides evidence that the
solution represents the tachyon vacuum in bosonic SFT.
Recently, a homotopy operator for the BRST operator
 was obtained \cite{Inatomi:2011xr} and 
it was applied to show moduli-independence of one loop vacuum energy at
the solution.

It is natural to ask how the identity-based solutions in bosonic SFT
can be extended to solutions in super SFT.
Marginal solutions given in \cite{Takahashi:2002ez} were already
extended to solutions in 
superstring field theory in \cite{Kishimoto:2005bs, Kishimoto:2005wv}.
Here, we will extend scalar solutions given in \cite{Takahashi:2002ez} 
to solutions in modified cubic superstring field theory
\cite{Preitschopf:1989fc, Arefeva:1989cp, Arefeva:1989cm}.

Actually, apart from the context of SFT,
we begin with a construction of deformed BRST operators
using integration of the BRST current in superstring theory
with some weighting function.
To find a nilpotent operator, it is necessary to introduce two other
operators in terms of ghost fields.
The resulting nilpotent deformed BRST operator has a similar form
to that of the theory expanded around the identity-based solution in
bosonic SFT. Using a method in \cite{Inatomi:2011xr}, we will show that
there exists a homotopy operator for a particular weighting function,
which is associated with the tachyon vacuum solution in bosonic SFT.

The situation with respect to the deformed BRST operators 
is very similar to the bosonic SFT.
We expect that it should correspond
to some classical solutions in super SFT in some sense.
Indeed, we will construct a class of identity-based solutions using
half-integration of the BRST current in modified cubic superstring field
theory.
In the theory around the solution, the deformed BRST operator
with the same function is reproduced.
The existence of a homotopy operator shows that the solution is
nontrivial although it can be rewritten as a pure gauge form at least
formally.
However, its physical interpretation is unclear so far
because the obtained solution is in the theory on a BPS D-brane.

This paper is organized as follows.
In the next section, we will construct a class of deformed BRST
operators. In \S \ref{sec:homotopy}, we will find their homotopy operators
in the case of particular functions and discuss the relationship to the
conventional BRST cohomology.
In \S \ref{sec:SSFT}, we will construct corresponding solutions in cubic
super SFT. Finally, we will give concluding remarks in \S \ref{sec:Rem}.
Some technical details for computations are given 
in Appendix \ref{sec:ComRel} and we will present
a summary of the BRST cohomology in Neveu-Schwarz (NS) $0$-picture and 
Ramond (R) $(-1/2)$-picture in Appendix \ref{sec:RNScoh}.

\section{Deformed BRST operators
\label{sec:deformed}}

We consider a deformation of the BRST operator in superstring theory
using $b$, $c$ and $\beta$, $\gamma$ ghosts:
\begin{eqnarray}
 Q'=Q(f)+C(g)+\Theta(h),
\end{eqnarray}
where each operator is defined by an integration along the unit circle
$|z|=1$,
\begin{eqnarray}
\label{eq:Qf}
 Q(f)&=&\oint \frac{dz}{2\pi i}\,f(z)\,j_{\rm B}(z),\\
\label{eq:Cg}
 C(g)&=&\oint \frac{dz}{2\pi i}\,g(z)\,c(z),\\
\label{eq:Th}
 \Theta(h)&=&\oint \frac{dz}{2\pi i}\,h(z)\,\theta(z).
\end{eqnarray}
Here,
$f(z)$, $g(z)$ and $h(z)$ are some weighting functions, and 
$j_{\rm B}(z)$ and $c(z)$ are the BRST current and the ghost field.
$\theta(z)$ is defined by
\begin{eqnarray}
 \theta(z)=c\beta \gamma(z)-\partial c(z),
\end{eqnarray}
which is a primary field with dimension 0.

First, we will show that if we impose nilpotency on the operator $Q'$,
the functions $g(z)$ and $h(z)$ can be related to $f(z)$, and
the resulting nilpotent operator is given by
\begin{eqnarray}
\label{eq:brs}
 Q'=Q(e^\lambda)+C\left(-\frac{1}{2}
(\partial \lambda)^2e^\lambda\right)
+\Theta\left(-\frac{1}{4}\partial e^\lambda\right).
\end{eqnarray}

To find the nilpotent deformed BRST operator,
it is necessary to calculate the operator product expansion (OPE) among
$j_{\rm B}(z)$, $c(z)$ and $\theta(z)$.
The BRST current is defined by
\begin{eqnarray}
 j_{\rm B}(z)&=& c T^{\rm m}(z)+\gamma G^{\rm m}(z)\nn
&&+bc\partial c(z)+\frac{1}{4}c\partial \beta \gamma(z)
-\frac{3}{4}c\beta \partial \gamma(z)
+\frac{3}{4}\partial c \beta \gamma (z)-b\gamma^2(z)
+\frac{3}{4}\partial^2 c(z),
\end{eqnarray}
where $T^{\rm m}(z)$ is the energy momentum tensor and $G^{\rm m}(z)$
is the supercurrent for the matter sector. 
Using the definition, the OPE of the BRST currents is
obtained as
\begin{eqnarray}
 j_{\rm B}(y)\,j_{\rm B}(z)&\sim&\frac{1}{(y-z)^3}\left\{
\left(\frac{43}{8}-\frac{c^{\rm m}}{2}\right)c\partial c(z)
+\left(\frac{2c^{\rm m}}{3}-7\right)\gamma^2(z)\right\}\nn
&&
+\frac{1}{(y-z)^2}
\frac{1}{2}\partial \left\{
\left(\frac{43}{8}-\frac{c^{\rm m}}{2}\right)c\partial c(z)
+\left(\frac{2c^{\rm m}}{3}-7\right)\gamma^2(z)\right\}\nn
&&
+\frac{1}{y-z}\left\{
\left(\frac{5}{4}-\frac{c^{\rm m}}{12}\right)c\partial^3 c(z)
+\left(\frac{c^{\rm m}}{3}-5\right)\gamma \partial^2\gamma(z)\right.\nn
&&
\label{eq:jbjb}
\left.
\ \ \ +\partial\left(
\frac{1}{4}c\gamma G^{\rm m}(z)+\frac{1}{2}bc\gamma^2(z)
+\frac{1}{4}\beta \gamma^3(z)\right)\right\},
\end{eqnarray}
where the matter central charge $c^{\rm m}$ is $15$.
Then, the OPEs of $j_{\rm B}(z)$, $\theta(z)$ and $c(z)$ are calculated as
\begin{eqnarray}
\label{eq:jbcbg}
 j_{\rm B}(y)\,\theta(z)&\sim&
\frac{1}{(y-z)^2}\left(
\frac{1}{4}c\partial c(z)-\gamma^2(z)\right)\nn
&&+\frac{1}{y-z}
\left(-c\gamma G^{\rm m}(z)-2bc\gamma^2 (z)
-\beta \gamma^3(z)\right),\\
\label{eq:jbc}
j_{\rm B}(y)\,c(z)&\sim& \frac{1}{y-z}(c\partial c(z)-\gamma^2(z)),\\
\label{eq:cbgcbg}
\theta(y)\,\theta(z)&\sim&
\frac{1}{y-z}c\partial c(z).
\end{eqnarray}

If the functions $f(z)$, $g(z)$ and $h(z)$ are holomorphic in
an annulus including the unit circle,
anti-commutation relations among
the operators (\ref{eq:Qf}), (\ref{eq:Cg}) and (\ref{eq:Th}) can be
derived from use of the OPEs~(\ref{eq:jbjb}), (\ref{eq:jbcbg}), 
(\ref{eq:jbc}) and (\ref{eq:cbgcbg}):
\begin{eqnarray}
 \{Q(f),\,Q(f)\}&=&\oint\frac{dz}{2\pi i}\,
\frac{-1}{2}(\partial f(z))^2\left(
-\frac{17}{8}c\partial c(z)+3\gamma^2(z)\right)\nn
\label{eq:QfQf}
&&+\oint\frac{dz}{2\pi i}\frac{-1}{4}\partial f^2(z)
\left(c\gamma G^{\rm m}(z)+2bc\gamma^2(z)+\beta \gamma^3(z)\right),\\
\label{eq:QfCg}
\{Q(f),\,C(g)\}&=& \oint \frac{dz}{2\pi i}
f(z)g(z)(c\partial c(z)-\gamma^2(z)),\\
\{Q(f),\,\Theta(h)\}
&=&
\oint\frac{dz}{2\pi i}\,
\frac{1}{4}h(z)\partial f(z)\,c\partial c(z)\nn
&&
-\oint\frac{dz}{2\pi i}\,
h(z)\partial f(z)\,\gamma^2(z)\nn
\label{eq:QfTh}
&&-\oint\frac{dz}{2\pi i}f(z)h(z)
\left(c\gamma G^{\rm m}(z)+2bc\gamma^2(z)+\beta \gamma^3(z)\right),\\
\label{eq:ThTh}
\{\Theta(h),\,\Theta(h)\}
&=&
\oint \frac{dz}{2\pi i}h^2(z)c\partial c(z).
\end{eqnarray}
The same anti-commutation relations also hold for the functions defined
only on the unit circle.
As in the bosonic case~\cite{Takahashi:2002ez,Kishimoto:2002xi},
these can be obtained by Fourier analysis with the use of the
commutation relations of oscillators in Appendix \ref{sec:ComRel}.

Using (\ref{eq:QfQf}), (\ref{eq:QfCg}), (\ref{eq:QfTh}) and
(\ref{eq:ThTh}),
we can find the anti-commutator of the deformed BRST operator,
\begin{eqnarray}
 \{Q',\,Q'\}&=&
\oint\frac{dz}{2\pi i}
\left\{
\frac{17}{16}(\partial f)^2+h^2+2fg+
\frac{1}{2}h\partial f\right\}
c\partial c(z)\nn
&&
+\oint\frac{dz}{2\pi i}
\left\{
-\frac{3}{2}(\partial f)^2-2fg-2h\partial f
\right\}
\gamma^2(z)\nn
&&
+\oint\frac{dz}{2\pi i}
\frac{-1}{4}\left\{
\partial f^2 +8 fh
\right\}
\left(
c\gamma G^{\rm m}(z)+2bc\gamma^2(z)+\beta\gamma^3(z)\right).
\end{eqnarray}
For nilpotency of $Q'$, the functions
$f(z)$, $g(z)$ and $h(z)$ should satisfy the following equations:
\begin{eqnarray}
&&
\label{eq:siki1}
 \frac{17}{16}(\partial f)^2+h^2+2fg+
\frac{1}{2}h\partial f=0,\\
&&
\label{eq:siki2}
-\frac{3}{2}(\partial f)^2-2fg-2h\partial f=0,\\
&&
\label{eq:siki3}
\partial f^2 +8 fh=0.
\end{eqnarray}
{}From these equations, we find that the functions $g(z)$ and $h(z)$
are given in terms of $f(z)$,
\begin{eqnarray}
 g(z)=-\frac{(\partial f)^2}{2f},
\ \ \ \ 
 h(z)=-\frac{1}{4}\partial f(z).
\end{eqnarray}
Writing $f(z)=e^{\lambda(z)}$, we can find that
the nilpotent deformed BRST operator is given by the equation
(\ref{eq:brs}).

Next, we will show that the nilpotent operator $Q'$
can be expressed as a similarity
transformation constructed by the ghost number current
$j_{\rm gh}(z)=-bc(z)-\beta\gamma(z)$.
We can find the following OPEs of $j_{\rm gh}(z)$:
\begin{eqnarray}
 j_{\rm gh}(y)\,j_{\rm B}(z)&\sim&
\frac{2}{(y-z)^3}c(z)
+\frac{1}{(y-z)^2}\left(\partial c(z)
-\frac{1}{4}\theta(z)\right)+\frac{1}{y-z}
j_{\rm B}(z),
\label{eq:ope_jgh1}
\\
j_{\rm gh}(y)\,c(z)&\sim& \frac{1}{y-z}c(z),
\label{eq:ope_jgh2}
\\
j_{\rm gh}(y)\,\theta(z)&\sim& 
\frac{1}{y-z}\,\theta(z).
\label{eq:ope_jgh3}
\end{eqnarray}
Here, we introduce an integrated operator for
the ghost number current:
\begin{eqnarray}
 q(\lambda)&=& \oint \frac{dz}{2\pi i} \lambda(z)\,j_{\rm gh}(z),
\end{eqnarray}
where $\lambda(z)$ is a weighting function.
Similar to (\ref{eq:QfQf}), (\ref{eq:QfCg}), (\ref{eq:QfTh}) and
(\ref{eq:ThTh}), we can find the following commutation-relations:
\begin{eqnarray}
\label{eq:qQ}
 [q(\lambda),\,Q(f)]&=&Q(\lambda f)
+C\left(-\partial
f\,\partial \lambda\right)
+\Theta\left(-\frac{1}{4}f
\partial \lambda \right),\\
\label{eq:qC}
\left[q(\lambda),\,C(g)\right]&=&C(\lambda g),\\
\label{eq:qT}
\left[q(\lambda),\,\Theta(h)\right]&=&\Theta(h\lambda).
\end{eqnarray}

Let us define $Q_t$ with a parameter $t$:
\begin{eqnarray}
 Q_t &=& e^{t q(\lambda)}Q_{\rm B} e^{-t q(\lambda)}.
\label{eq:defQt}
\end{eqnarray}
Using the commutation relations (\ref{eq:qQ}), (\ref{eq:qC}) and
(\ref{eq:qT}), we can easily find that the operator $Q_t$
is expressed as
\begin{eqnarray}
\label{eq:Qt2}
 Q_t&=&Q(f_t)+C(g_t)+\Theta(h_t).
\end{eqnarray}
where $f_t(z),g_t(z)$ and $h_t(z)$ are functions with the parameter $t$.
Differentiating both sides of (\ref{eq:defQt}) with respect to $t$,
we have 
\begin{eqnarray}
 \frac{d}{dt}Q_t&=& 
Q(\lambda f_t)+C\left(-
\partial f_t\,\partial \lambda\right)
+\Theta\left(-\frac{1}{4}f_t\partial \lambda\right)\nn
&&+C(\lambda g_t)+\Theta(h_t\lambda),
\label{eq:dQt}
\end{eqnarray}
where we have used the expression (\ref{eq:Qt2}) and
the commutation relations
(\ref{eq:qQ}), (\ref{eq:qC}) and (\ref{eq:qT}).
Comparing (\ref{eq:dQt}) with the derivative of (\ref{eq:Qt2}),
we obtain the following differential equations:
\begin{eqnarray}
 \frac{d}{dt}f_t&=&
\lambda f_t, \\
\frac{d}{dt}g_t &=&
 -\partial f_t\,\partial \lambda
+\lambda g_t,\\
\frac{d}{dt}h_t&=& -\frac{1}{4}f_t\partial \lambda
+h_t\lambda.
\end{eqnarray}
We can easily solve these equations under the initial conditions,
$f_{t=0}=1,g_{t=0}=0,h_{t=0}=0$.
Finally, setting $t=1$, we find that the nilpotent BRST operator
(\ref{eq:brs}) is written as a similarity transformation:
\begin{eqnarray}
 Q'&=& e^{q(\lambda)}Q_{\rm B}e^{-q(\lambda)}.
\label{eq:simqQ-q}
\end{eqnarray}

\section{Homotopy operators in superstring theory
\label{sec:homotopy}
}

Here, we construct a homotopy operator for the deformed BRST operator
constructed in \S \ref{sec:deformed}
with a particular function $\lambda$.

Firstly, we note that the following OPEs:
\begin{eqnarray}
 j_{\rm B}(y)b(z)&\sim&\frac{3/2}{(y-z)^3}+\frac{1}{(y-z)^2}
\left(-bc(z)-\frac{3}{4}\beta\gamma(z)\right)
+\frac{1}{y-z}T(z),\\
c(y)b(z)&\sim&\frac{1}{y-z},\\
\theta(y)b(z)&\sim&\frac{1}{(y-z)^2}+\frac{1}{y-z}\beta\gamma(z),
\end{eqnarray}
which lead to following anti-commutation relations:
\begin{eqnarray}
\{Q(f),b(z)\}&=&\frac{3}{4}\partial^2f(z)+\partial
 f(z)\left(-bc(z)-\frac{3}{4}\beta\gamma(z)\right)
+f(z)T(z),\\
\{C(g),b(z)\}&=&g(z),\\
\{\Theta(h),b(z)\}&=&\partial h(z)+h(z)\beta\gamma(z).
\end{eqnarray}
Hence, anti-commutator of $Q'$ (\ref{eq:brs}) and $b(z)$ is calculated as
\begin{eqnarray}
 \{Q',b(z)\}&=&\frac{1}{2}(\partial^2\lambda(z))e^{\lambda(z)}
+(\partial\lambda(z))e^{\lambda(z)}j_{\rm gh}(z)
+e^{\lambda(z)}T(z).
\label{eq:Qpbz}
\end{eqnarray}
If we choose the function $\lambda(z)$ such that $e^{\lambda(z)}$ 
has a second order zero $z_0$, the above relation implies that
$\{Q',b(z)\}$ becomes a c-number at $z=z_0$.
For example, let us take the function $\lambda=h^l_a$:
\begin{eqnarray}
&&h^l_a(z)=\log\left(1-\frac{a}{2}(-1)^l(z^l-(-1)^lz^{-l})^2\right),
~~~~~(a\ge -1/2;~l=1,2,3,\cdots)
\label{eq:h_adef}
\end{eqnarray}
which was adopted in \cite{Takahashi:2002ez,Kishimoto:2002xi}
for identity-based solutions in bosonic SFT.
Although $e^{h^l_a}$ does not have second order zeros
in the case $a>-1/2$, 
$e^{h^l_{a=-1/2}}$ has second order zeros at 
$z_k=e^{i\frac{k-1}{l}\pi}$ for odd $l$ and 
$z_k=e^{i\frac{2k-1}{2l}\pi}$ for even $l$ ($k=1,2,\cdots,2l$),
which are solutions to $z^{2l}+(-1)^l=0$.
Namely, the deformed BRST operator $Q'$ with the function
$\lambda=h^l_{a=-1/2}$ has a homotopy operator $\hat A$, such as
\begin{eqnarray}
 \{Q',\hat A\}=1,~~~~~\hat A^2=0.
\label{eq:homQA}
\end{eqnarray}
In this case, the operator $\hat A$, which is
BPZ even and Hermitian operator, is explicitly, given by
\begin{eqnarray}
 \hat A&=&\sum_{k=1}^{2l}a_kl^{-2}z_k^2\,b(z_k),
~~~~\sum_{k=1}^{2l}a_k=1,
\label{eq:hatAex}
\end{eqnarray}
where
\begin{eqnarray}
 a_k=a_{l-k+2},~~(k=1,2,\cdots,l+1);~~~~~
a_k=a_{3l-k+2},~~(k=l+2,l+3,\cdots,2l)
\label{eq:BPZlodd}
\end{eqnarray}
for odd $l$,
\begin{eqnarray}
 a_k=a_{l-k+1},~~(k=1,2,\cdots,l);~~~~~
a_k=a_{3l-k+1},~~(k=l+1,l+2,\cdots,2l)
\label{eq:BPZleven}
\end{eqnarray}
for even $l$ and $a_k\in \mathbb{R}\,(k=1,2,\cdots, 2l)$.
The above $\hat{A}$ is exactly the same form as a bosonic counterpart in
\cite{Inatomi:2011xr}.
In the same way, using the commutation relations obtained by differentiating 
(\ref{eq:Qpbz}), homotopy operators for the function such that
$e^{\lambda}$ has higher-order zeros
investigated in \cite{Igarashi:2005wh}
can be also constructed as in the bosonic case \cite{Inatomi:2011xr}.

The existence of a homotopy operator $\hat A$ such as (\ref{eq:homQA})
implies that the deformed BRST operator $Q'$ has vanishing cohomology:
\begin{eqnarray}
 Q'\psi=0 ~~~&\Leftrightarrow&~~~\psi=Q'(\hat A\psi).
\label{eq:trivialcohomology}
\end{eqnarray}
On the other hand, $Q'$ can be expressed as a similarity transformation
of the conventional $Q_{\rm B}$: 
$Q'=e^{q(h^l_{-1/2})}Q_{\rm B}e^{-q(h^l_{-1/2})}$ from (\ref{eq:simqQ-q})
and, at least formally, we have
\begin{eqnarray}
Q_{\rm B}\Psi=0~~~\Leftrightarrow~~~Q'(e^{q(h^l_{-1/2})}\Psi)=0
~~~\Leftrightarrow~~~
e^{q(h^l_{-1/2})}\Psi=Q'(\hat{A}e^{q(h^l_{-1/2})}\Psi).
\label{eq:Formal}
\end{eqnarray}
Therefore, the nontrivial part of $Q_{\rm B}$-cohomology also becomes trivial
in terms of $Q'$-cohomology by 
multiplying $e^{q(h^l_{-1/2})}$.
To find out what is happening, let us more concretely see products of
$e^{q(h^l_{-1/2})}$ and the $Q_{\rm B}$-closed states
 in $0$-picture (\S\ref{sec:NS0coh}) and $(-1/2)$-picture
 (\S\ref{sec:R-1/2coh}), which
are pictures for the NS and R sector, respectively, in modified cubic
superstring field theory.
We decompose $q(h^l_{-1/2})$ to positive, zero and negative mode part:
\begin{eqnarray}
&&q(h^l_{-1/2})=q^{(+)}(h^l_{-1/2})+q^{(0)}(h^l_{-1/2})+q^{(-)}(h^l_{-1/2}),
\end{eqnarray}
where
\begin{eqnarray}
&&q^{(0)}(h^l_{-1/2})=-q_0\log 4,~~~~~~
q^{(\pm)}(h^l_{-1/2})=-\sum_{n=1}^{\infty}\frac{(-1)^{n(l+1)}}{n}q_{\pm
 2nl},
\end{eqnarray}
for $j_{\rm gh}(z)=\sum_nq_nz^{-n-1}$ and (\ref{eq:h_adef}).
Noting that $[q_n,q_m]=0$ and $q_0$ counts the ghost number, we find
\begin{eqnarray}
e^{q(h^l_{-1/2})}\Psi&=&
U2^{-2}\mathcal{P}|{\rm tach}\rangle_0+
U2^{-4}\mathcal{P}'\biggl(c_0|{\rm tach}\rangle_0
+\frac{\sqrt{2}}{\sqrt{\alpha'}k^+}\gamma_{-\frac{1}{2}}|0,k_1\rangle_0
\biggr)\nonumber\\
&&+Q'(e^{q(h^l_{-1/2})}\chi),
\label{eq:UNS}
\end{eqnarray}
for $Q_{\rm B}$-closed $\Psi$ in the NS $0$-picture with $p^+\ne 0$ from
(\ref{eq:Th_0}), where $U$ is given by negative modes of $j_{\rm gh}$:
\begin{eqnarray}
 U=\exp\left(q^{(-)}(h^l_{-1/2})\right)
=\exp\left(-\sum_{n=1}^{\infty}\frac{(-1)^{n(l+1)}}{n}q_{-2nl}\right).
\label{eq:Usoldef}
\end{eqnarray}
Using the above $U$, we have
\begin{eqnarray}
e^{q(h^l_{-1/2})}\Psi&=&
U2^{-1}|{\cal P}\rangle_{-\frac{1}{2}}
+U2^{-3}(c_0+\gamma_0\vartheta)|{\cal P}'\rangle_{-\frac{1}{2}}
+Q'(e^{q(h^l_{-1/2})}\chi),
\label{eq:UR}
\end{eqnarray}
for $Q_{\rm B}$-closed $\Psi$ in the R $(-1/2)$-picture with $p^+\ne 0$ from
(\ref{eq:Th_-1/2}).
Furthermore, for the sake of completeness, we also mention the exceptional
case, namely zero momentum sector, as follows:
\begin{eqnarray}
e^{q(h^l_{-1/2})}\Psi&=&U{\cal C}^{(0)}
b_{-1}\left|\downarrow\right\rangle_0
+U2^{-2}{\cal C}^{(1)}_{\mu}
(\alpha_{-1}^{\mu}+\psi^{\mu}_{-\frac{1}{2}}b_{-1}\gamma_{\frac{1}{2}})
\left|\downarrow\right\rangle_0
\nonumber\\
&&+U2^{-4}{\cal C}^{(2)}_{\mu}
\left(\alpha_{-1}^{\mu}c_0+2\psi_{-\frac{1}{2}}^{\mu}\gamma_{-\frac{1}{2}}
+\psi_{-\frac{1}{2}}^{\mu}b_{-1}c_0\gamma_{\frac{1}{2}}
\right)
\left|\downarrow\right\rangle_0\nonumber\\
&&+U2^{-6}{\cal C}^{(3)}\left(
2\gamma_{-\frac{1}{2}}^2
+c_{-1}c_0
+\gamma_{-\frac{1}{2}}\gamma_{\frac{1}{2}}
b_{-1}c_0
\right)\left|\downarrow\right\rangle_0+Q'(e^{q(h^l_{-1/2})}\chi),~~~
\label{eq:UNSp0}
\end{eqnarray}
for $Q_{\rm B}$-closed $\Psi$ in the NS $0$-picture from (\ref{eq:Th_0_0})
and
\begin{eqnarray}
e^{q(h^l_{-1/2})}\Psi
&=&
U2^{-1}A_0^a|S_a\rangle_{-\frac{1}{2}}
+U2^{-3}A_1^a\gamma_0|S_a\rangle_{-\frac{1}{2}}
+Q'(e^{q(h^l_{-1/2})}\chi),
\label{eq:URp0}
\end{eqnarray}
for $Q_{\rm B}$-closed $\Psi$ 
in the R $(-1/2)$-picture from (\ref{eq:Th_-1/2_0}).

In (\ref{eq:UNS}), (\ref{eq:UR}), (\ref{eq:UNSp0}) and (\ref{eq:URp0}),
the same $U$ (\ref{eq:Usoldef}) is multiplied on the nontrivial part
of $Q_{\rm B}$-cohomology. By multiplying the homotopy operator $\hat{A}$
(\ref{eq:hatAex}) and move it to the right, we have
\begin{eqnarray}
&&\hat{A}U(\cdots)
=\exp\left(-\sum_{n=1}^{\infty}\frac{1}{n}\right)U\hat{A}(\cdots)
=0,
\label{eq:AU0UA}
\end{eqnarray}
where we have used $[q_n,b_m]=-b_{n+m}$, which leads to
$U^{-1}b(z)U
=e^{-\sum_{n=1}^{\infty}\frac{(-1)^{n(l+1)}}{n}z^{-2nl}}b(z)$.
The relation (\ref{eq:AU0UA}) implies that, for the nontrivial part
$|\varphi\rangle$ in $Q_{\rm B}$-cohomology in the NS and R sector, all
coefficients of $\hat{A}e^{q(h^l_{-1/2})}|\varphi\rangle$ vanish in the
Fock space.
On the other hand, the last equation in (\ref{eq:Formal}) indicates the
relation $|\varphi\rangle=Q'\hat{A}e^{q(h^l_{-1/2})}|\varphi\rangle$,
which may be interpreted that $|\varphi\rangle$ is $Q'$-exact outside
the Fock space.
Here, we regard a space in which states are expressed 
as a linear combination of states made of $b_{-n},c_{-m}$ ($n\ge
2,m\ge -1$) on the conformal vacuum $|0\rangle_{bc}$ in the $bc$ ghost
sector and $\beta_{-r},\gamma_{-s}$ $(r\ge P+3/2,s\ge -P-1/2)$ 
on $|P\rangle_{\beta\gamma}$
in the $\beta\gamma$ ghost sector in $P$-picture
as ``the Fock space.''
Although we should define it as a completion with respect to 
an appropriate norm mathematically, we leave it ambiguous at this
stage.
Intuitively, we could interpret that $\hat{A} e^{q(h^l_{-1/2})}\varphi$
weakly converges to zero but not in a strong sense.

For comparison, let us comment on a similar situation on
the identity-based solution with $h^l_a$ in the bosonic SFT investigated
in \cite{Inatomi:2011xr}. 
The nontrivial part $|\varphi\rangle$ 
of $Q'$-cohomology in terms of the Fock space
found in \cite{Kishimoto:2002xi} can be regarded as $Q'$-exact outside
the Fock space by the homotopy operator $\hat{A}$.
However, the similarity transformation from the Kato-Ogawa BRST
operator: $Q'=e^{q(h^l_a)}Q_{\rm B}e^{-q(h^l_a)}$ becomes ill-defined at
$a=-1/2$ in the bosonic case.
In fact, $U_l$ included in $|\varphi\rangle$ \cite{Inatomi:2011xr,
Kishimoto:2002xi} is different from the above $U$ (\ref{eq:Usoldef}).


\section{Classical solutions in modified cubic superstring field
 theory
\label{sec:SSFT}}

In \S\ref{sec:deformed}, we have constructed a class of deformed BRST
operators, which are nilpotent.
In \S\ref{sec:homotopy}, we have constructed homotopy operators at the
boundary of the parameter $a$
included in the associated functions $h_a^l$.
We expect that they may be realized as BRST operators
at appropriate classical solutions in superstring field theory.
In this section, we show that it is indeed the case 
in the framework of modified cubic superstring field theory.

Using the analogy of a class of identity-based solutions
\cite{Takahashi:2002ez} in bosonic string field theory, 
we expect that a string field,
which is given by 
\begin{eqnarray}
  A_c&=&Q_L(f)I + C_L(g)I
+\Theta_L(h)I
\label{eq:sol_SSFT}
\end{eqnarray}
with appropriate functions: $f,g$ and $h$,
may be a classical solution in superstring
field theory.
Here, we have used half integrations:
\begin{eqnarray}
&&Q_L(f)=\!\int_{C_{\rm L}}\!\frac{dz}{2\pi i}f(z)j_{\rm B}(z),~~
C_L(g)=\!\int_{C_{\rm L}}\!\frac{dz}{2\pi i}g(z)c(z),~~
\Theta_L(h)=\!\int_{C_{\rm L}}\!\frac{dz}{2\pi i}h(z)\theta(z),
~~~~~~
\label{eq:halfint1}
\end{eqnarray}
where the subscript $C_{\rm L}$ for each integral
denotes a half unit circle corresponding to
a left half of an open string.
$I$ denotes the identity state, given by 
the total (matter$+$ghost) Virasoro generators corresponding to a
conformal map $2z/(1-z^2)$
in the same way as bosonic open SFT.

The string field $A_c$ (\ref{eq:sol_SSFT})
has the ghost number one and the picture number zero
in the NS sector.
Therefore, it is natural to treat it in the framework of 
modified cubic superstring field theory, whose action is
\begin{eqnarray}
 S[A,\Psi]&=&\frac{1}{2}\langle A,Y_{-2}Q_{\rm B}A\rangle
+\frac{1}{3}\langle A,Y_{-2}A*A\rangle
+\frac{1}{2}\langle\Psi,YQ_{\rm B}\Psi\rangle
+\langle A,Y\Psi*\Psi\rangle.
\label{eq:action_s}
\end{eqnarray}
In the above, $A$ and $\Psi$ are string fields in the NS and R
sector, respectively.
$Y_{-2}$ and $Y$ are inverse picture changing operators with 
the picture number $-2$ and $-1$, respectively, inserted at the
midpoint.
$\langle\,\cdot\,,\,\cdot\,\rangle$ denotes the BPZ inner product in the
small Hilbert space.
(Here, we use the $(\beta,\gamma)$ system instead of
$(\xi,\eta,\phi)$-system for the superghost sector
in terms of the worldsheet CFT.)
The equations of motion for the action (\ref{eq:action_s}) are
\begin{eqnarray}
 &&Y_{-2}(Q_{\rm B}A+A*A)+Y\Psi*\Psi=0,
~~~~~Y(Q_{\rm B}\Psi+A*\Psi+\Psi*A)=0.
\label{eq:EOMs}
\end{eqnarray}
In the following, we will find the string field
$A_c$ (\ref{eq:sol_SSFT}) in the NS sector, which satisfies
\begin{eqnarray}
 Q_{\rm B}A_c+A_c*A_c=0,
\label{eq:EOM_bos}
\end{eqnarray}
by choosing appropriate functions: $f,g$ and $h$.
Our strategy is almost the same as the bosonic case elaborated
in \cite{Takahashi:2002ez}. 
For a primary field $\sigma$ with conformal dimension $h$, 
we have a relation from the definition of the star
product of string fields in open SFT with a midpoint interaction:
\begin{eqnarray}
&&(\Sigma_R(F)B_1)*B_2=-(-1)^{|\sigma||B_1|}B_1*(\Sigma_L(F)B_2),
\label{eq:Sigma_RL}
\end{eqnarray}
where $\Sigma_L(F)$ and $\Sigma_R(F)$ are half integrations of a primary
field $\sigma(z)$ with the conformal dimension $h$
multiplied by a function $F(z)$ such as $F(-1/z)=(z^2)^{1-h}F(z)$:
\begin{eqnarray}
&&\Sigma_L(F)=\int_{C_{\rm L}}\frac{dz}{2\pi i}F(z)\sigma(z),
~~~~
\Sigma_R(F)=\int_{C_{\rm R}}\frac{dz}{2\pi i}F(z)\sigma(z),
\label{eq:Sigma_LR}
\end{eqnarray}
and $B_1,B_2$ are any string fields.
$C_{\rm L}$ ($C_{\rm R}$) denotes a half unit circle with positive
(negative) real part. $(-1)^{|\sigma||B_1|}$ in (\ref{eq:Sigma_RL}) is 
$-1$ in the case that both $\sigma(z)$ and $B_1$  are Grassmann odd
and $+1$ otherwise.
In the case that $B_1$ and/or $B_2$ are the identity state, 
which is the identity element with respect to the star product,
we have
\begin{eqnarray}
 &&\Sigma_R(F)I=-\Sigma_L(F)I,
\label{eq:Sigma_RLI}
\\
&&(\Sigma_L(F)I)*B=\Sigma_L(F)B,
~~~B*(\Sigma_L(F)I)=-(-1)^{|\sigma||B|}\Sigma_R(F)B.
\label{eq:Sigma_LILI}
\end{eqnarray}
Assuming relations of functions:
\begin{eqnarray}
 f(-1/z)=f(z),~~~g(-1/z)=z^4\,g(z),~~~~h(-1/z)=z^2\,h(z),
\label{eq:fgh}
\end{eqnarray}
and noting that $j_{\rm B}(z),c(z)$ and $\theta(z)$ are primary fields with 
conformal dimension $1,-1$ and $0$, respectively,
the half integrations included in the string field (\ref{eq:sol_SSFT})
are the same form as in (\ref{eq:Sigma_LR}).
Therefore, we can use the formula (\ref{eq:Sigma_LILI}) and
$A_c*A_c$ is computed as
\begin{eqnarray}
A_c*A_c&=&
\biggl[\frac{1}{2}\{Q_L(f),Q_L(f)\}+
\frac{1}{2}\{C_L(g),C_L(g)\}+
\frac{1}{2}\{\Theta_L(h),\Theta_L(h)\}
\nonumber\\
&&+\{Q_L(f),C_L(g)\}+\{Q_L(f),\Theta_L(h)\}
+\{C_L(g),\Theta_L(h)\}
\biggr]I
\nonumber\\
&=&\biggl[
-\frac{7}{32}\{\Theta_L(\partial f),\Theta_L(\partial f)\}
+\frac{3}{4}\{Q_{\rm B},C_L((\partial f)^2)\}+\frac{1}{4}\{Q_{\rm B},
\Theta_L(f\partial f)\}
\nonumber\\
&&+\frac{1}{2}\{\Theta_L(h),\Theta_L(h)\}
+\{Q_{\rm B},C_L(fg)\}+\{Q_{\rm B},\Theta_L(fh)\}
\nonumber\\
&&-\frac{3}{4}\{\Theta_L(\partial f),\Theta_L(h)\}
+\{Q_{\rm B},C_L((\partial f)h)\}
\biggr]I.
\label{eq:Ac*Ac}
\end{eqnarray}
In the second equality, we have used 
some relations:
(\ref{eq:QLQL}), (\ref{eq:QLCL}) and (\ref{eq:QLThetaL})
described in Appendix \ref{sec:ComRel} and 
 $f(\pm i)=0$ is supposed in order to perform partial integrations.
Using $Q_{\rm B}I=0$, (\ref{eq:QBQL})
and (\ref{eq:Ac*Ac}), 
the left hand side of (\ref{eq:EOM_bos}) is
\begin{eqnarray}
&&Q_{\rm B}A_c+A_c*A_c\nonumber\\
&&=\biggl[\Bigl\{Q_{\rm B},C_L\Bigl((1+f)g+\frac{3}{4}(\partial f)^2+h\partial
 f\Bigr)\Bigr\}+
\Bigl\{Q_{\rm B}, \Theta_L\Bigl(
(1+f)\Bigl(h+\frac{1}{4}\partial f\Bigr)
\Bigr)
\Bigr\}
\nonumber\\
&&~~~~~~
-\frac{7}{32}\{\Theta_L(\partial f),\Theta_L(\partial f)\}
+\frac{1}{2}\{\Theta_L(h),\Theta_L(h)\}
-\frac{3}{4}\{\Theta_L(\partial f),\Theta_L(h)\}
\biggr]I.
\end{eqnarray}
Therefore, introducing a function $\lambda$, we have
\begin{eqnarray}
&&f=e^{\lambda}-1,~~~~
g=-\frac{1}{2}(\partial\lambda)^2e^{\lambda},
~~~~
h=-\frac{1}{4}(\partial\lambda)e^{\lambda},
\label{eq:fghlambda}
\end{eqnarray}
by imposing (\ref{eq:EOM_bos}).
The function $\lambda$ should satisfy
\begin{eqnarray}
&&\lambda(-1/z)=\lambda(z),~~~~\lambda(\pm i)=0,
\label{eq:lambda_cond}
\end{eqnarray}
in order to guarantee the assumptions for functions
(\ref{eq:fgh}) and $f(\pm i)=0$.

Therefore, the relation (\ref{eq:EOM_bos}) implies that
the string field $A_c$ (\ref{eq:sol_SSFT}) with (\ref{eq:fghlambda}) 
and (\ref{eq:lambda_cond}) in the NS sector 
(and vanishing string field in the R sector)
can be regarded as a classical solution to the
equations of motion (\ref{eq:EOMs}).
The action around the solution is obtained by re-expanding
(\ref{eq:action_s}) (and subtracting the $S[A_c,0]$) 
as
\begin{eqnarray}
 S'[A,\Psi]&\equiv&S[A+A_c,\Psi]-S[A_c,0]
\nonumber\\
&=&\frac{1}{2}\langle A,Y_{-2}Q'A\rangle
+\frac{1}{3}\langle
A,Y_{-2}A*A\rangle+\frac{1}{2}\langle\Psi,YQ'\Psi\rangle
+\langle A,Y\Psi*\Psi\rangle,
\label{eq:action_sp}
\end{eqnarray}
where the BRST operator $Q'$ at the solution is defined by
\begin{eqnarray}
 Q'B&=&Q_{\rm B}B+A_c*B-(-1)^{|B|}B*A_c
\end{eqnarray}
for a string field $B$. ($(-1)^{|B|}$ denotes 
the Grassmann parity of the string field $B$.)
Substituting the concrete expression of $A_c$
to the above formula
and using the relations (\ref{eq:Sigma_LILI}), 
the BRST operator $Q'$ at the solution $A_c$ is obtained:
\begin{eqnarray}
 Q'&=&Q_{\rm B}+(Q_L(f)+C_L(g)+\Theta_L(h))+
(Q_R(f)+C_R(g)+\Theta_R(h))
\nonumber\\
&=&Q(e^{\lambda}) + C\left(-\frac{1}{2}
(\partial\lambda)^2e^{\lambda}\right)
+\Theta\left(-\frac{1}{4}(\partial\lambda)e^{\lambda}\right),
\label{eq:QpSFT}
\end{eqnarray}
where the last expression is given by integrations along 
a full circle.
It is equal to the deformed BRST operator (\ref{eq:brs}) 
investigated in \S\ref{sec:deformed}.

Next, let us consider the relation to the 
expression using a similarity transformation:
$Q'=e^{q(\lambda)}Q_{\rm B}e^{-q(\lambda)}$ (\ref{eq:simqQ-q}).
As in the case of bosonic SFT, one expects that $A_c$ may be
related to
$e^{q_L(\lambda)I}Q_{\rm B}e^{-q_L(\lambda)I}$,
where the symbol ``$*$'' for the star product among string fields 
is omitted. 
For the half integrations of $j_{\rm gh}$:
\begin{eqnarray}
&&q_L(\lambda)=\!\int_{C_{\rm L}}\!\frac{dz}{2\pi i}\lambda(z)j_{\rm gh}(z),
~~~
q_R(\lambda)=\!\int_{C_{\rm R}}\!\frac{dz}{2\pi i}\lambda(z)j_{\rm gh}(z),
\label{eq:halfint2}
\end{eqnarray}
we note that (\ref{eq:Sigma_RL}) should be modified such as
\begin{eqnarray}
&&(q_R(\lambda)B_1)*B_2=-B_1*(q_L(\lambda)B_2)
+k(\lambda)B_1*B_2,~~~~
k(\lambda)=\int_{C_{\rm L}}\frac{dz}{2\pi i}\frac{\lambda(z)}{z},~~~
\end{eqnarray}
for $\lambda(-1/z)=\lambda(z)$
because the ghost number current $j_{\rm gh}=-bc-\beta\gamma$ is
not primary and is transformed as
$j_{\rm gh}(z)=w^2\,j_{\rm gh}(w)-w$ for $w=-1/z$.
Then, corresponding to (\ref{eq:Sigma_RLI}) and (\ref{eq:Sigma_LILI}),
we have the relations:
\begin{eqnarray}
 &&q_R(\lambda)I=-q_L(\lambda)I+k(\lambda)I,
~~~~~~
(q_L(\lambda)I)*B=q_L(\lambda)B.
\end{eqnarray}
Using the second equation in the above and
the commutation relations given in (\ref{eq:qLrel0}),
(\ref{eq:qLrel1}), (\ref{eq:qLrel2})
and (\ref{eq:qLrel3}),  we can calculate as follows:
\begin{eqnarray}
e^{q_L(\lambda)I}Q_{\rm B}e^{-q_L(\lambda)I}&=&
(e^{q_L(\lambda)}Q_{\rm B}e^{-q_L(\lambda)})I
 =\left(
\sum_{k=1}^{\infty}\frac{1}{k!}\,({\rm ad}_{q_L(\lambda)})^kQ_{\rm B}
\right)I
\nonumber\\
&=&\left(
Q_L\left(e^{\lambda}-1\right)
-\frac{1}{2}C_L\left((\partial\lambda)^2e^{\lambda}\right)
-\frac{1}{4}\Theta_L\left((\partial\lambda)e^{\lambda}\right)
\right)I,
\label{eq:puregauge}
\end{eqnarray}
for a function $\lambda$ such as (\ref{eq:lambda_cond}).
(We have denoted as ${\rm ad}_XY\equiv [X,Y]$.)
The last expression in (\ref{eq:puregauge})
is equal to $A_c$ (\ref{eq:sol_SSFT}) with (\ref{eq:fghlambda}).
Namely, the identity-based solution $A_c$ in the NS sector,
constructed in this section, can be rewritten as
a pure gauge form.
Here we note that the action (\ref{eq:action_s})
is invariant under a part of gauge transformations of string fields
in the NS and R sector:
\begin{eqnarray}
&&A'=e^{-\Lambda}A\,e^{\Lambda}+e^{-\Lambda}Q_{\rm B}e^{\Lambda},
~~~~\Psi'=e^{-\Lambda}\Psi\,e^{\Lambda},
\end{eqnarray}
with a gauge parameter string field $\Lambda$, which 
is Grassmann even and has ghost number zero and picture number zero.

One may think that the solution $A_c$ is trivial from the relation
(\ref{eq:puregauge}).
However, the action around the solution given in (\ref{eq:action_sp}) 
is nontrivially different from the original one (\ref{eq:action_s})  
for a particular type of the function $\lambda$, for
example, $\lambda=h^l_{a=-1/2}$  (\ref{eq:h_adef}),
because the BRST operator $Q'$ (\ref{eq:QpSFT}) becomes cohomologically
trivial due to the existence of a homotopy operator 
$\hat A$ (\ref{eq:homQA})
as was investigated in \S\ref{sec:homotopy}.

In the case of bosonic SFT, for $\lambda=h^l_{a=-1/2}$, 
the pure gauge form of the identity-based solution
 becomes singular \cite{Takahashi:2002ez} because 
the operator $e^{\pm q_L(\lambda)}$ becomes ill-defined
due to the divergent factor in the normal ordered expression.
It is caused by the singular OPE of the ghost number current:
$j_{\rm gh}(y)j_{\rm gh}(z)\sim (y-z)^{-2}$.
On the other hand, in the case of superstring field theory, 
$j_{\rm gh}(y)j_{\rm gh}(z)$ $(y\to z)$ is regular
because the ghost number current
included in (\ref{eq:puregauge})
is  $j_{\rm gh}=-bc-\beta\gamma$ instead of $-bc$.
Therefore, the pure gauge form,
which is the first expression in (\ref{eq:puregauge}),
is not singular even for $\lambda=h^l_{a=-1/2}$ in this sense.
If we regard the pure gauge form in (\ref{eq:puregauge})
as well-defined, we can interpret that
the solution $A_c$ in modified cubic superstring field
theory corresponds to the solution: $\Phi_c=-q_L(\lambda)I$
in Berkovits' WZW-like superstring field theory \cite{Berkovits:1995ab}
because $A_c$ is in the small Hilbert space and the equation of motion
$\eta_0(e^{-\Phi_c}Q_{\rm B}e^{\Phi_c})=0$ is satisfied. 
In this case, it is also pure gauge because it can be rewritten as 
$\Phi_c=\eta_0(-\xi_0q_L(\lambda)I)$ in the large Hilbert space.

\section{Concluding remarks
\label{sec:Rem}
}

In this paper, we have constructed a class of deformed BRST operators
$Q'$, which is nilpotent and has ghost number one in the context of RNS
formalism of open superstring.
The operator $Q'$ is given by integrations with
the BRST current, $c$ ghost, $\theta\equiv c\beta\gamma-\partial c$
and  an appropriate function $\lambda$.
In the case of a particular type function $\lambda$, we have found
a homotopy operator for $Q'$ such as $\{Q',\hat A\}=1$, 
which implies that the cohomology of $Q'$ vanishes.

In the framework of modified cubic superstring field theory, we have
constructed an identity-based solution $A_c$ in the NS sector,
in the theory around which the BRST operator coincides with the deformed
BRST operator, i.e., $Q'=Q_{\rm B}+[A_c,\,\cdot\,\}_*$.
Therefore, corresponding to a particular type of the function such as
$\lambda=h^l_{a=-1/2}$, the solution seems to be nontrivial
and the vanishing cohomology of $Q'$ might imply that the D-brane
vanishes as in the case of bosonic SFT.

The deformed BRST operator $Q'$ can be re-expressed as a
similarity transform of the conventional BRST operator $Q_{\rm B}$ 
using the integration of the ghost number current 
$j_{\rm gh}=-bc-\beta\gamma$
with the function $\lambda$: $Q'=e^{q(\lambda)}Q_{\rm B}e^{-q(\lambda)}$.
For a particular function, such as $\lambda=h^l_{a=-1/2}$,
the normal ordered form of $e^{\pm q(\lambda)}$ is ill-defined in the
bosonic case, but it is not so in the superstring. Nevertheless, a
homotopy operator can be found, which may imply that the corresponding
identity-based solution in cubic superstring field theory,
investigated in \S \ref{sec:SSFT}, would be nontrivial although
it has a pure gauge expression (\ref{eq:puregauge}) formally.
In fact, the image of nontrivial part $|\varphi\rangle$ of the
$Q_{\rm B}$-cohomology: $e^{q(h^l_{-1/2})}|\varphi\rangle$
is mapped to the states outside the Fock space by the homotopy
operator $\hat{A}$ in the sense that all coefficients of 
$\chi'\equiv\hat{A}e^{q(h^l_{-1/2})}|\varphi\rangle$ vanish 
in the Fock space and $e^{q(h^l_{-1/2})}|\varphi\rangle
=Q'\chi'$ holds 
as far as we respect the relation  $\{Q',\hat A\}=1$.
It is necessary to define space of states or string fields more
rigorously in order to clarify this delicate issue.

The identity-based solution, which we have constructed, is a universal
solution in the cubic superstring field theory.
The solution might represent a vacuum where a D-brane vanishes because
the cohomology of the BRST operator becomes empty.
However, this interpretation might be obscure because 
the original theory is on a BPS D-brane, which should be stable.
The situation may be similar to the Erler solution \cite{Erler:2007xt},
at which there exists a homotopy operator and its energy reproduces a
D-brane tension.

To confirm the non-triviality of the obtained solution
for particular functions, such as $\lambda=h^l_{a=-1/2}$,
the evaluations of the vacuum energy or the gauge
invariant overlap\footnote{
See \cite{Ellwood:2008jh, Kawano:2008ry} for details.
} are desired.
However, direct computations are difficult as in the case of bosonic SFT
because $A_c$ is an identity-based solution.
Namely, indefinite quantities will appear because the identity state
has vanishing width in terms of the sliver frame.
To avoid this singularity, we can map the solution of 
$Q_{\rm B}A+A*A=0$ to other solution, which is given by a linear
combination of states with finite
width, as was elaborated in \cite{Kishimoto:2007bb},
for example.
In principle, we can evaluate the vacuum energy or the gauge invariant
overlap for the mapped solution as a regularization.
In the action of modified cubic superstring field theory, the inverse
picture changing operators, $Y$ and $Y_{-2}$, are inserted at the
midpoint, which might cause another problem.
However, we find that the OPEs: $j_{\rm B}(y)Y(z)$, $c(y)Y(z)$,
$\theta(y)Y(z)$ ($y\to z$) are regular,
where $Y=Y(i)=c(i)\delta'(\gamma(i))$ and $Y_{-2}=Y(i)Y(-i)$
in terms of the $(\beta,\gamma)$ system.
 Therefore, there are no apparent
divergences caused by the collisions of $Y,Y_{-2}$ and the operators on
the identity state in the solution $A_c$.

\section*{Acknowledgments}
I.~K. and T.~T. would like to thank Yuji Igarashi and Katsumi Itoh for
collaborations in 2005 at the very early stage of the present work.
I.~K. would like to thank Maiko Kohriki and Hiroshi Kunitomo for helpful
comments.
The authors thank the Yukawa Institute for Theoretical Physics at Kyoto
University. Discussions during the YITP workshop YITP-W-11-05 on ``Field
Theory and String Theory'' were useful to complete this work.
The work of I.~K. and T.~T. is supported by
JSPS Grant-in-Aid for Scientific Research (C) (\#21540269).

\appendix

\section{Commutation relations
\label{sec:ComRel}}

In this appendix, we collect some formulae related to (anti-)commutation
relations.

We expand $j_{\rm B},c,\theta,j_{\rm gh}$ as
$j_{\rm B}(z)=\sum_nQ_nz^{-n-1},c(z)=\sum_nc_nz^{-n+1},\theta(z)=\sum_n\theta_nz^{-n}$
as usual.
The anti-commutation relations among them can be derived from their
OPEs, (\ref{eq:jbjb}), (\ref{eq:jbcbg}), (\ref{eq:jbc}) and
(\ref{eq:cbgcbg}):
\begin{eqnarray}
\{Q_n,Q_m\}&=&
-\frac{nm}{2}\oint\frac{dz}{2\pi i}z^{n+m-2}
\left(-\frac{17}{8}c\partial c(z)+3\gamma^2(z)\right)
\nonumber\\
&&+\frac{n+m}{4}\oint\frac{dz}{2\pi i}z^{n+m-1}\left(
-c\gamma G^{\rm m}(z)-2bc\gamma^2(z)-\beta\gamma^3(z)\right),\\
\{Q_n,c_m\}&=&\oint\frac{dz}{2\pi i}z^{n+m-2}\left(c\partial c(z)
-\gamma^2(z)\right),\\
\{Q_n,\theta_m\}&=&n\oint\frac{dz}{2\pi i}z^{n+m-2}
\left(\frac{1}{4}c\partial c(z)-\gamma^2(z)\right)
\nonumber\\
&&+\oint \frac{dz}{2\pi i}
z^{n+m-1}\left(
-c\gamma G^{\rm m}(z)-2bc\gamma^2(z)-\beta\gamma^3(z)\right),\\
\{\theta_n,\theta_m\}&=&\oint\frac{dz}{2\pi i}z^{n+m-2}c\partial c(z),
\end{eqnarray}
which imply following relations
\begin{eqnarray}
\{Q_n,Q_m\}&=&nm\left(-\frac{7}{16}\{\theta_n,\theta_m\}
+\frac{3}{2}\{Q_{\rm B},c_{n+m}\}\right)
+\frac{n+m}{4}\{Q_{\rm B},\theta_{n+m}\},
\label{eq:Qn_rel1}
\\
\{Q_n,c_m\}&=&\{Q_{\rm B},c_{n+m}\},
\label{eq:Qn_rel2}\\
\{Q_n,\theta_m\}&=&\{Q_{\rm B},\theta_{n+m}\}
+n\left(-\frac{3}{4}\{\theta_n,\theta_m\}+\{Q_{\rm B},c_{n+m}\}
\right).
\label{eq:Qn_rel3}
\end{eqnarray}
Similarly, by expanding the ghost number current as
$j_{\rm gh}(z)=\sum_nq_nz^{-n-1}$, we find the commutation relations:
\begin{eqnarray}
&&[q_n,Q_m]=-mnc_{n+m}+Q_{n+m}-\frac{n}{4}\theta_{n+m},
~~~[q_n,c_m]=c_{n+m},~~~[q_n,\theta_m]=\theta_{n+m},
~~~~~
\label{eq:qn_com}
\end{eqnarray}
{}from the OPEs given in (\ref{eq:ope_jgh1}), (\ref{eq:ope_jgh2}) and
(\ref{eq:ope_jgh3}).
In the context of a construction of identity-based solutions 
in superstring filed theory, it is necessary to compute
(anti-)commutation relations including half integrations:
(\ref{eq:halfint1}) and (\ref{eq:halfint2}).
Using the above relations: (\ref{eq:Qn_rel1}),
(\ref{eq:Qn_rel2}), (\ref{eq:Qn_rel3}) and (\ref{eq:qn_com}),
and the splitting properties of the delta function,
$\delta(z,w)=\sum_nz^{n-1}w^{-n}$,
clarified in \cite{Takahashi:2002ez}:
\begin{eqnarray}
\int_{C_{\rm L}}\frac{dz}{2\pi i}
\int_{C_{\rm L}}\frac{dw}{2\pi i}
F(z)G(w)\delta(z,w)=
\int_{C_{\rm L}}\frac{dz}{2\pi i}F(z)G(z),
\end{eqnarray}
we have
\begin{eqnarray}
 \{Q_{\rm B},Q_L(f)\}&=&\frac{1}{4}\{Q_{\rm B},\Theta_L(\partial f)\},
\label{eq:QBQL}
\\
 \{Q_L(f),Q_L(f)\}&=&-\frac{7}{16}\{\Theta_L(\partial
 f),\Theta_L(\partial f)\}
+\frac{3}{2}\{Q_{\rm B},C_L((\partial f)^2)\}
\nonumber\\
&&+\frac{1}{2}\{Q_{\rm B},\Theta_L(f\partial f)\},
\label{eq:QLQL}
\\
\{Q_L(f),C_L(g)\}&=&\{Q_{\rm B},C_L(fg)\},
\label{eq:QLCL}
\\
\{Q_L(f),\Theta_L(h)\}&=&\{Q_{\rm B},\Theta_L(fh)\}-\frac{3}{4}
\{\Theta_L(\partial f),\Theta_L(h)\}
+\{Q_{\rm B},C_L((\partial f)h)\},~~~~
\label{eq:QLThetaL}
\\
{}[q_L(\lambda),Q_{\rm B}]&=&
-\frac{1}{4}\Theta_L(\partial\lambda)+Q_L(\lambda),
\label{eq:qLrel0}\\
{}[q_L(\lambda),Q_L(\Lambda)]&=&-C_L(\partial\lambda\partial\Lambda)
-\frac{1}{4}\Theta_L((\partial\lambda)\Lambda)+Q_L(\lambda\Lambda),
\label{eq:qLrel1}\\
{}[q_L(\lambda),C_L(\Lambda)]&=&C_L(\lambda\Lambda),
\label{eq:qLrel2}\\
{}[q_L(\lambda),\Theta_L(\Lambda)]&=&\Theta_L(\lambda\Lambda).
\label{eq:qLrel3}
\end{eqnarray}
Here, functions: $f,\lambda$ and $\Lambda$ should vanish at 
the points: $z=\pm i$, which are the boundaries of the half unit circle,
 in the above formulae including $\partial
 f,\partial\lambda,\partial\Lambda$ to perform partial integrations.

\section{On the BRST cohomology for RNS string
\label{sec:RNScoh}}

In this appendix, we summarize the results of the conventional
$Q_{\rm B}$-cohomology in the Fock space without $b_0$ (and $\beta_0$) gauge
condition in $0$-picture and $(-1/2)$-picture with a brief outline of
proofs.
In \S\ref{sec:homotopy}, we discuss the relation between nontrivial part
of the $Q_{\rm B}$-cohomology and the homotopy operator for $Q'$.
Here, we note that
\begin{eqnarray}
&&\beta_r|P\rangle_{\beta\gamma}=0~~(r>-P-\frac{3}{2}),
~~~~
\gamma_r|P\rangle_{\beta\gamma}=0~~(r>P+\frac{1}{2}),
\end{eqnarray}
in the $\beta\gamma$-sector in $P$-picture.

\subsection {Neveu-Schwarz $0$-picture
\label{sec:NS0coh}}

Let us review the $Q_{\rm B}$-cohomology in the NS sector in terms of
$0$-picture.
When we restrict the Fock space to ${\rm Ker}\,b_0$ and $p^+\ne 0$,
nontrivial part of the cohomology can be written using the DDF
operators in the matter sector. Namely, we have \cite{Kohriki:2010ry}
\begin{eqnarray}
Q_{\rm B}\Psi=0,~b_0\Psi=0&\Leftrightarrow&
\Psi={\cal P}|{\rm tach}\rangle_0+Q_{\rm B}\chi,
\label{eq:KKM}
\end{eqnarray}
where $\chi\in {\rm Ker}\,b_0$, ${\cal P}$ is made of the DDF operators,
which (anti-)commute with $Q_{\rm B}$,
 and $|{\rm tach}\rangle_0$ is the onshell tachyon in the $0$-picture:
\begin{eqnarray}
&&|{\rm tach}\rangle_0=
\left(\psi_{-\frac{1}{2}}^--\frac{1}{\sqrt{2\alpha'}k^+}
b_{-1}\gamma_{\frac{1}{2}}
+\frac{1}{4\alpha'(k^+)^2}\psi_{-\frac{1}{2}}^+
\right)|0,k_1\rangle_0,\\
&&|0,k_1\rangle_0=|k_1\rangle_{\rm mat}\otimes c_1|0\rangle_{bc}\otimes
 |P=0\rangle_{\beta\gamma},~~~~k_1^+=k^+,~k_1^-=\frac{1}{4\alpha'k^+},~k^i=0.
\end{eqnarray}
Using this result,\footnote{
We can apply a similar method to \cite{Henneaux:1986kp} 
for bosonic string, with a slight modification caused by $M|{\rm
tach}\rangle_0\ne 0$.
In the case of $(-1)$-picture, where 
the physical tachyon $|0,k_1\rangle_{-1}
=\delta(\gamma_{\frac{1}{2}})|0,k_1\rangle_0$
satisfies $M|0,k_1\rangle_{-1}=0$ as in bosonic string, the result is
\cite{Lian:1989cy}:
\begin{eqnarray}
&&Q_{\rm B}\Psi=0~~\Leftrightarrow~~\Psi
=\mathcal{P}|0,k_1\rangle_{-1}+
c_0\mathcal{P}'|0,k_1\rangle_{-1}+Q_{\rm B}\chi.
\label{eq:Th_-1}
\end{eqnarray}
Multiplying the picture changing operator 
$X(z)=\{Q_{\rm B},\Theta(\beta(z))\}=G(z)\delta(\beta(z))-\partial
b(z)\delta'(\beta(z))$
and taking the limit $z\to 0$, we can also rederive
the relation (\ref{eq:Th_0}) in 0-picture.
} \ 
we can demonstrate
\begin{eqnarray}
 &&Q_{\rm B}\Psi=0~~\Leftrightarrow~~\Psi
=\mathcal{P}|{\rm tach}\rangle_0+
\mathcal{P}'\biggl(c_0|{\rm tach}\rangle_0
+\frac{\sqrt{2}}{\sqrt{\alpha'}k^+}\gamma_{-\frac{1}{2}}|0,k_1\rangle_0
\biggr)
+Q_{\rm B}\chi,~~~~
\label{eq:Th_0}
\end{eqnarray}
for the Fock space without $b_0$-gauge condition.
Here, $\mathcal{P}$ and $\mathcal{P}'$ are generated by the DDF operators.

To derive  (\ref{eq:Th_0}), 
we note the similarity transformed expression
of the BRST operator obtained in \cite{Kohriki:2011zza}:
\begin{eqnarray}
&&Q_{\rm B}=e^{-R_2-R_3}(c_0L+\sqrt{2\alpha'}A)e^{R_2+R_3},
\label{eq:KKM_sim}
\end{eqnarray}
where $L=\{b_0,Q_{\rm B}\}$,
\begin{eqnarray}
 A&=&p^+\left(\sum_{n\neq 0}c_{-n}\alpha_n^{-}+
\sum_{r\in\mathbb{Z}+\frac{1}{2}}\gamma_{-r}\psi_r^{-}\right),
\label{eq:Adef}
\end{eqnarray}
and
\begin{eqnarray}
R_2&=&\frac{1}{\sqrt{2\alpha'}\,p^+}
\Biggl[\,
\sum_{\substack{m\ne 0,n\ne 0,\\m+n\ne 0}}
\left(
\frac{m+n}{2mn}\alpha_{-m}^+\alpha_{-n}^+\alpha_{m+n}^-
+\frac{n}{m}\alpha_{-m}^+c_{-n}b_{m+n}
\right)
\nonumber\\
&&~~~~+\sum_{\substack{m,n,\\m\ne
 0}}\frac{1}{2m}\alpha_{-m}^+\alpha_{-n}^i\alpha_{m+n}^i
+\sum_{\substack{n\in\mathbb{Z},\\
r\in\mathbb{Z}+\frac{1}{2}
}}\psi^+_{-r}\alpha_{-n}^i\psi_{n+r}^i
\nonumber\\
&&~~~~+\sum_{\substack{n\ne 0,\\
r\in\mathbb{Z}+\frac{1}{2}
}}\Biggl\{
\left(\frac{3}{2}+\frac{r}{n}\right)\alpha_{-n}^+\psi_{-r}^+\psi_{n+r}^-
+\left(\frac{1}{4}+\frac{r}{2n}\right)\alpha_{-n}^+\psi^i_{-r}\psi^i_{n+r}
\nonumber\\
&&~~~~~~~~~~~~~
-\left(\frac{1}{2}+\frac{r}{n}\right)\alpha_{-n}^+\gamma_{-r}\beta_{n+r}
+n\psi^+_{-r}c_{-n}\beta_{n+r}-\psi_{-r}^+b_{-n}\gamma_{n+r}
\Biggr\}
\Biggr],
\label{eq:R2def}
\\
R_3&=&\frac{1}{\sqrt{2\alpha'}\,p^+}b_0
\left(\sum_{n\neq 0}c_{-n}\alpha_n^{+}+
\sum_{r\in\mathbb{Z}+\frac{1}{2}}\gamma_{-r}\psi_r^{+}\right).
\label{eq:R3def}
\end{eqnarray}
Then, we find the following relations:
\begin{eqnarray}
&&e^{-R_2-R_3}c_0e^{R_2+R_3}=
c_0-[R_3,c_0],
\label{eq:c0_sim}
\\
&&Q_{\rm B}(c_0-[R_3,c_0])|{\rm tach}\rangle_0=0,
\label{eq:QBclosed0}\\
&&-[R_3,c_0]|{\rm tach}\rangle_0=
\frac{\sqrt{2}}{\sqrt{\alpha'}k^+}\gamma_{-\frac{1}{2}}|0,k_1\rangle_0
+Q_{\rm B}\biggl(\frac{-1}{2\alpha'(k^+)^2}
\psi^+_{-\frac{1}{2}}|0,k_1\rangle_0\biggr).
\label{eq:QBequiv0}
\end{eqnarray}
Firstly, we apply (\ref{eq:KKM}) for the sector projected by $c_0b_0$
and then, using (\ref{eq:KKM}), (\ref{eq:QBclosed0}) and
(\ref{eq:QBequiv0}) for ${\rm Ker}\,b_0$, we can conclude
(\ref{eq:Th_0}).\\

In the case of zero momentum states, we cannot apply (\ref{eq:Th_0})
due to $p^+=0$.
However, for $L\ne 0$ sector, the $Q_{\rm B}$-cohomology is
trivial because of $\{Q_{\rm B},b_0/L\}=1$.
Hence, we investigate the $Q_{\rm B}$-cohomology in ${\rm Ker}\,L$.
Furthermore, we consider cohomology for each ghost number sector,
where its dimension is finite although there is a
creation operator $\gamma_{\frac{1}{2}}$ with negative level:
$-\frac{1}{2}$, in $0$-picture unlike the case of $(-1)$-picture.\footnote{
In  $(-1)$-picture, the $Q_{\rm B}$-cohomology can be easily
investigated by considering all ghost number sector in ${\rm Ker}\,L$
with zero momentum. The result is \cite{Lian:1989cy}:
\begin{eqnarray}
&&Q_{\rm B}\Psi=0\Leftrightarrow
\Psi={\cal
C}^{(0)}\beta_{-\frac{1}{2}}\left|\downarrow\right\rangle_{-1}
+{\cal
C}^{(1)}_{\mu}\psi^{\mu}_{-\frac{1}{2}}\left|\downarrow\right\rangle_{-1}+
{\cal
C}^{(2)}_{\mu}\psi^{\mu}_{-\frac{1}{2}}c_0\left|\downarrow\right\rangle_{-1}
+{\cal
C}^{(3)}\gamma_{-\frac{1}{2}}c_0\left|\downarrow\right\rangle_{-1}+Q_{\rm B}\chi,
~~~~~~~~
\label{eq:coh-1_0}
\end{eqnarray}
where ${\cal
C}^{(0)},{\cal
C}^{(1)}_{\mu},{\cal
C}^{(2)}_{\mu}$ and ${\cal C}^{(3)}$ are constants and
$\left|\downarrow\right\rangle_{-1}=|0\rangle_{\rm mat}\otimes 
c_1|0\rangle_{bc}\otimes
|P=-1\rangle_{\beta\gamma}$.
By multiplying the picture changing operator $X(z)$ to
(\ref{eq:coh-1_0}) and taking the limit $z\to 0$, we can also rederive
(\ref{eq:Th_0_0}) by subtracting $Q_{\rm B}$-exact states.
} 
For $\Psi=|\psi\rangle+c_0|\phi\rangle$ ($|\psi\rangle,|\phi\rangle\in
{\rm Ker}\,L\cap{\rm Ker}\,b_0$), 
$Q_{\rm B}\Psi=0$ imposes $\widetilde{Q}|\phi\rangle=0$, where
\begin{eqnarray}
 Q_{\rm B}=c_0L+b_0M+\widetilde{Q}.
\end{eqnarray}
Furthermore, $\widetilde{Q}$, which does not include $b_0$ and $c_0$,
can be expanded with respect to $\gamma_{\frac{1}{2}}$ and 
its canonical conjugate $\beta_{-\frac{1}{2}}$ as 
\begin{eqnarray}
&&\widetilde{Q}=\gamma_{\frac{1}{2}}\widehat{G}_{-\frac{1}{2}}
-b_{-1}\gamma_{\frac{1}{2}}^2+\widehat{K}_{\frac{1}{2}}\beta_{-\frac{1}{2}}
+\widehat{Q}.
\end{eqnarray}
Noting that there exists a non-negative integer $N$ for a
finite ghost number sector, $|\phi\rangle$
can be expanded as
\begin{eqnarray}
&&|\phi\rangle=\sum_{k=0}^N\gamma_{\frac{1}{2}}^k|\phi_k\rangle,
~~~~\beta_{-\frac{1}{2}}|\phi_k\rangle=0,
\end{eqnarray}
and the condition $\widetilde{Q}|\phi\rangle=0$ imposes
$b_{-1}|\phi_{N}\rangle=0$, which implies
$|\phi_{N}\rangle=b_{-1}|\phi'_{N}\rangle$
$(\exists\,|\phi_N'\rangle\in {\rm Ker}\,b_0\cap
      {\rm Ker}\,\beta_{-\frac{1}{2}})$.
Then, $|\phi\rangle+Q_{\rm B}(\gamma_{\frac{1}{2}}^{N-2}|\phi_N'\rangle)$
cancels the $O(\gamma_{\frac{1}{2}}^N)$-term for $N-2\ge 0$.
Repeating this procedure, $|\phi\rangle$ can be rewritten as
\begin{eqnarray}
&&|\phi\rangle=|\phi_0'\rangle+\gamma_{\frac{1}{2}}|\phi_1'\rangle+Q_{\rm
 B}\chi_{\phi},
~~~~|\phi_0'\rangle,|\phi_1'\rangle\in {\rm Ker}\,b_0\cap {\rm
Ker}\,\beta_{-{\frac{1}{2}}},~~
\chi_{\phi}\in{\rm Ker}\,b_0,~~~
\end{eqnarray}
with $L|\phi_0'\rangle=0$ and $L|\phi_1'\rangle=\frac{1}{2}|\phi_1'\rangle$.
Using the above also for $|\psi\rangle$, we find all solutions to
$Q_{\rm B}\Psi=0$ up to $Q_{\rm B}$-exact terms. The result
for zero momentum sector is
\begin{eqnarray}
&&Q_{\rm B}\Psi=0~~~\Leftrightarrow~~\nonumber\\
&&\Psi={\cal C}^{(0)}
b_{-1}\left|\downarrow\right\rangle_0
\!+\!{\cal C}^{(1)}_{\mu}
(\alpha_{-1}^{\mu}\!+\!\psi^{\mu}_{-\frac{1}{2}}b_{-1}\gamma_{\frac{1}{2}})
\left|\downarrow\right\rangle_0
\!+\!{\cal C}^{(2)}_{\mu}\!
\left(\!\alpha_{-1}^{\mu}c_0\!+\!2\psi_{-\frac{1}{2}}^{\mu}\gamma_{-\frac{1}{2}}
\!+\!\psi_{-\frac{1}{2}}^{\mu}b_{-1}c_0\gamma_{\frac{1}{2}}
\!\right)\!
\left|\downarrow\right\rangle_0\nonumber\\
&&~~~~~~+{\cal C}^{(3)}\left(
2\gamma_{-\frac{1}{2}}^2
+c_{-1}c_0
+\gamma_{-\frac{1}{2}}\gamma_{\frac{1}{2}}
b_{-1}c_0
\right)\left|\downarrow\right\rangle_0+Q_{\rm B}\chi,
\label{eq:Th_0_0}
\end{eqnarray}
where ${\cal C}^{(0)},{\cal C}^{(1)}_{\mu},{\cal C}^{(2)}_{\mu}$
and ${\cal C}^{(3)}$ are constants
and $\left|\downarrow\right\rangle_0=|0\rangle_{\rm mat}\otimes 
c_1|0\rangle_{bc}\otimes
|P=0\rangle_{\beta\gamma}$.

\subsection{Ramond $(-1/2)$-picture
\label{sec:R-1/2coh}}

Here, we review the $Q_{\rm B}$-cohomology in the R sector with
$(-1/2)$-picture \cite{Henneaux:1987ux,Lian:1989cy}.
Firstly, we note that the $L(\equiv\{Q_{\rm B},b_0\})\ne 0$ sector is
$Q_{\rm B}$-trivial also in the R sector.
By restricting the Fock space to ${\rm Ker}\,b_0\cap {\rm Ker}\,
\beta_0$ and $p^+\ne 0$, 
the $Q_{\rm B}$-cohomology was investigated in \cite{Ito:1985qa} and the
result is
\begin{eqnarray}
Q_{\rm B}\Psi=0,~b_0\Psi=0,~\beta_0\Psi=0&\Leftrightarrow&
\Psi=|{\cal P}\rangle_{-\frac{1}{2}}+Q_{\rm B}\chi,
\label{eq:IMNU}
\end{eqnarray}
where $|{\cal P}\rangle_{-\frac{1}{2}}$ indicates
 the states generated by the
transverse operators, which (anti-)commute with $Q_{\rm B}$, in the matter
sector and $c_1|0\rangle_{bc}\otimes |P=-1/2\rangle_{\beta\gamma}$
in the ghost sector.
Let us remove the $b_0$ and $\beta_0$ gauge condition.
For $\Psi=|\psi\rangle+c_0|\phi\rangle$ ($|\psi\rangle,|\phi\rangle\in {\rm
Ker}\,L\cap{\rm Ker}\,b_0$), 
$Q_{\rm B}\Psi=0$ imposes $\widetilde{Q}|\phi\rangle=0$,
 where $\widetilde{Q}$ is also defined by $ Q_{\rm B}=c_0L+b_0M+\widetilde{Q}$
 in the R sector. 
Furthermore,
$\widetilde{Q}$, which does not include $b_0$ and
$c_0$, can be expanded with respect to $\gamma_0$ and
its canonical conjugate $\beta_0$ as 
\begin{eqnarray}
&&\widetilde{Q}=\gamma_0 F+\beta_0 K +\widehat{Q},\\
&&F=\sum_{n\in\mathbb{Z}}\alpha_{-m}\cdot\psi_m+\sum_{n\ne 0}\left(
\frac{n}{2}\beta_nc_{-n}-2b_{-n}\gamma_n\right).
\end{eqnarray}
We suppose that there exists a non-negative integer $N$ such that 
\begin{eqnarray}
&&|\phi\rangle=\sum_{k=0}^N\gamma_0^k|\phi_k\rangle,
~~~~\beta_0|\phi_k\rangle=0.
\end{eqnarray}
Then, the condition $\widetilde{Q}|\phi\rangle=0$ imposes
$F|\phi_{N}\rangle=0$, which implies
$|\phi_{N}\rangle=F|\phi'_{N}\rangle$
$(\exists\,|\phi_N'\rangle\in {\rm Ker}\,b_0\cap
      {\rm Ker}\,\beta_0)$,
because $F^2=L$ (i.e., $F$\,:\,nilpotent in ${\rm Ker}\,L$) and
$\{F,\vartheta\}=1$ with $\vartheta\equiv \psi_0^+/(\sqrt{2\alpha'}p^+)$.
Therefore, $|\phi\rangle-Q_{\rm B}(\gamma_0^{N-1}|\phi_N'\rangle)$
cancels the $O(\gamma_0^N)$-term for $N-1\ge 0$.
Repeating this procedure, $|\phi\rangle$ can be rewritten as
\begin{eqnarray}
&&|\phi\rangle=|\phi_0'\rangle+Q_{\rm B}\chi_{\phi},
~~~~|\phi_0'\rangle\in {\rm Ker}\,L\cap{\rm Ker}\,b_0\cap {\rm
Ker}\,\beta_0,~~
\chi_{\phi}\in{\rm Ker}\,b_0.
\end{eqnarray}
Using this fact and (\ref{eq:IMNU}), we have
\begin{eqnarray}
&&Q_{\rm B}\Psi=0~~\Rightarrow
~~~|\Psi\rangle=|\psi\rangle+c_0|{\cal P}\rangle_{-\frac{1}{2}}
+Q_{\rm B}\chi_{\phi},
\end{eqnarray}
by redefining $|\psi\rangle,\chi_{\phi}\in {\rm Ker}\,b_0$ appropriately,
where $|{\cal P}\rangle_{-\frac{1}{2}}$ denotes the transverse states.
As in the case of NS $0$-picture in \S\ref{sec:NS0coh}, we should note
that $M|{\cal P}\rangle_{-\frac{1}{2}}\ne 0$ and the similarity
transformed expression of $Q_{\rm B}$ (\ref{eq:KKM_sim}) holds for 
$A,R_2$ and $R_3$ by replacing $\sum_{r\in \mathbb{Z}+\frac{1}{2}}$ with 
$\sum_{r\in \mathbb{Z}}$ in (\ref{eq:Adef}), (\ref{eq:R2def}) and
(\ref{eq:R3def}).
Then, we find the relations: (\ref{eq:c0_sim}) and 
\begin{eqnarray}
&&Q_{\rm B}(c_0-[R_3,c_0])|{\cal P}\rangle_{-\frac{1}{2}}=0,
\\
&&-[R_3,c_0]|{\cal P}\rangle_{-\frac{1}{2}}
=\gamma_0\vartheta|{\cal P}\rangle_{-\frac{1}{2}}.
\end{eqnarray}
Using the above, we can show
\begin{eqnarray}
&&Q_{\rm B}\Psi=0~~~\Leftrightarrow
~~~\Psi=|{\cal P}\rangle_{-\frac{1}{2}}
+(c_0+\gamma_0\vartheta)|{\cal P}'\rangle_{-\frac{1}{2}}
+Q_{\rm B}\chi,
\label{eq:Th_-1/2}
\end{eqnarray}
where $|{\cal P}\rangle_{-\frac{1}{2}}$ and $|{\cal
P}'\rangle_{-\frac{1}{2}}$ are transverse states
in ${\rm Ker}\,L\cap{\rm Ker}\,b_0\cap {\rm
Ker}\,\beta_0$.\\

In the zero momentum sector, (\ref{eq:Th_-1/2}) cannot be applied
because $\vartheta=\psi_0^+/(\sqrt{2\alpha'}p^+)$ 
is not well-defined due to $p^+=0$.
However, noting that $M$ includes $-\gamma_0^2$
and ${\rm Ker}\,L$ is spanned by states of the form:
\begin{eqnarray}
&&\Psi=\sum_{k=0}^{\infty}A_k^a\gamma_0^k|S_a\rangle_{-\frac{1}{2}}
+c_0\sum_{k=0}^{\infty}B_k^a\gamma_0^k|S_a\rangle_{-\frac{1}{2}},
\label{eq:KerL_R}
\end{eqnarray}
where $|S_a\rangle_{-\frac{1}{2}}$ is a ground state with space-time
spinor index $a$ and $A_k^a,B_k^a$ are constants,
 we can demonstrate \cite{Lian:1989cy}:
\begin{eqnarray}
 Q_{\rm B}\Psi=0~~~\Leftrightarrow~~~
\Psi=A_0^a\,|S_a\rangle_{-\frac{1}{2}}
+A_1^a\,\gamma_0|S_a\rangle_{-\frac{1}{2}}+Q_{\rm B}\chi,
\label{eq:Th_-1/2_0}
\end{eqnarray}
in the zero momentum sector in $(-1/2)$-picture.

\end{document}